# *Direct Measurement of Storage and Loss Behavior in AFM Force-Distance Experiments Using the Modified Fourier Transformation*

*Berkin Uluutku, Marshall R. McCraw and Santiago D. Solares\**

*Department of Mechanical and Aerospace Engineering, The George Washington University School of Engineering and Applied Science, Washington, District of Columbia, USA*

*Corresponding Author: Santiago D. Solares – ssolares@gwu.edu

**ABSTRACT**

Force-distance curve experiments are commonly performed in Atomic Force Microscopy (AFM) to obtain the viscoelastic characteristics of materials, such as the storage and loss moduli or compliances. The classic methods used to obtain these characteristics consist of fitting a viscoelastic material model to the experimentally obtained AFM data. Here, we demonstrate a new method that utilizes the modified discrete Fourier transform to approximate the storage and loss behavior of a material directly from the data, without the need for a fit. Additionally, one may still fit a model to the resulting storage and loss behavior if a parameterized description of the material is desired. In contrast to fitting the data to a model chosen *a priori*, departing from a model-free description of the material's frequency behavior guides the selection of the model, such that the user may choose the one that is most appropriate for the particular material under study. To this end, we also include modified Fourier domain descriptions of commonly used viscoelastic models.

## I. INTRODUCTION

Atomic Force Microscopy (AFM) is widely utilized for micro and nanoscale viscoelastic characterization of biological samples and other soft materials. [1–5] A common technique used to perform this characterization relies on fitting a contact mechanics model to the force-indentation data obtained from the AFM experiment. There exist various types of models used to perform this analysis, including generalized spring-dashpot models, power-law rheology models, and fractional viscoelastic models; each providing a different approach with its own advantages and disadvantages.[5–7] Regardless of the chosen model, performing a fit to obtain viscoelastic properties has some caveats.

First, shortcomings in the fitting process may result in widely different combinations of model parameters that lead to similar force-indentation relationships for identical experimental data. Second, the viscoelastic properties that are obtained by performing the fit will be specific to the type of model chosen for the analysis, which may lead to certain physical phenomena not being captured by the selected model (e.g., a model that is unable to reproduce creep will not properly represent a viscoelastic material). Furthermore, in most cases, researchers are interested in frequency-dependent viscoelastic behaviors such as the storage and loss moduli or compliances. Since these properties are defined in the Fourier domain, obtaining them from fits of non-harmonic force-indentation experiments is only an indirect approach that may be prone to large errors in the frequency domain. Clearly, the most direct way of obtaining harmonic behavior would be using Fourier analysis; however, as the data from the force-distance experiments is numerically unbounded (i.e., it is aperiodic), it is not possible to directly employ a discrete Fourier analysis. [8]

In this paper, we present an analysis technique that utilizes the modified discrete Fourier transform to obtain an approximate representation of the storage and loss behavior of a material, directly from the experimental data. Although the technique does not rely on models, the user may still choose to fit the storage and loss behavior representations obtained from traditional viscoelastic models. Having a description of the material behavior prior to choosing a model provides very valuable guidance for selecting an appropriate material model. This work relies on a previously published mathematical development used to invert AFM force-distance data, which extends through a systematic treatment of the data that enables the use of the discrete Fourier transform.[9] We first introduce the theoretical background and transformation techniques used and then discuss the proposed method's application to experimental AFM data in detail. Common viscoelastic models and their definitions in the frequency domain are also provided for purposes of performing fits to force spectra. Additionally, we provide supplemental code and examples.[10,11]

## II. THEORETICAL BACKGROUND

The Laplace domain is central to many of the developments of the theory of linear viscoelasticity.[12–14] As here we seek to work with experimentally obtained data, we turn our focus to the discrete analog of the Laplace domain, namely the Z-domain. We can write the general stress-strain relationship for a linear viscoelastic material in the Z-domain as seen in Equations 1 and 2, in which $\sigma$ is the stress, $\epsilon$ is the strain, and $Q$ and $U$ are the source functions that define the material behavior, and are called the relaxance and retardance, respectively.[12,13] By separating the real and imaginary components of relaxance and retardance, one obtains the storage and loss behaviors, respectively.

$$\sigma(z) = Q(z)\epsilon(z) \tag{1}$$

$$\epsilon(z) = U(z)\sigma(z) \tag{2}$$

Since the stress and strain are not direct observables in typical AFM experiments, Equations 1 and 2 can be reformulated in terms of the AFM cantilever force and the probe indentation by using contact mechanics models. The Hertzian contact model can be considered to be the most well-known contact model, but it is only applicable for elastic materials.[15] For a viscoelastic material, the model of Lee and Radok is commonly utilized.[16–18] We can write the Lee and Radok model for a spherical indenter interacting with a flat surface in the Z-domain as seen in Equations 3 and 4, where $F$ is force, $h$ is indentation, $R$ is the indenter radius (AFM tip radius of curvature), and $Z\{h^{3/2}\}$ denotes the Z-transform (discrete analog of the Laplace transform) of the $\frac{3}{2}$ power of the indentation. For different geometries of indenters, we can write this relation in slightly different ways, which have been included in the supporting information.[2,19] We also would like to underline that, in agreement with the contact model, we exponentiate the indentation before taking the Z-transform.

$$Z\left\{h^{\frac{3}{2}}\right\} = \frac{3}{16\sqrt{R}} U(z)F(z) \qquad (3)$$

$$F(z) = \frac{16\sqrt{R}}{3} Q(z) Z\left\{h^{\frac{3}{2}}\right\} \qquad (4)$$

The Z-transform takes a discrete signal $X$ defined as a sequence in time and maps it onto the Z-domain, following the transformation described in Equation 5.[20,21] The domain variable $z$ is a complex number and is often represented in polar coordinates as $re^{i\omega}$. By expanding the definition of the Z-transform in terms of these polar coordinates, we arrive at the rightmost form of Equation 5, which shows the intimate relationship between the Z-transform and the discrete Fourier transform (DFT). More specifically, when the Z-transform is evaluated along the unit circle of the Z-domain (when $r$ is equal to 1) the evaluation is no different than the discrete Fourier transform. This is an established method to calculate the DFT and is called chirp Z-transform.[20] Likewise, each circle of constant radius (different from unity) represents a *modified* discrete Fourier transform (MDFT) of the original signal. One might notice that the MDFT of $X$ is equivalent to the discrete Fourier transform of the product of $X$ and the exponential term $r^{-n}$, where $n$ denotes the timestep index in the data sequence. In the introduction, it was stated that the Fourier transform cannot be directly used to analyze the force-indentation data, as they are numerically unbounded (aperiodic). However, the introduction of this exponential term allows for the discrete Fourier transform to be taken for certain values of $r$, for which the product is bounded within the window of measurement, as seen for the red trace ($r = 1.007$) in Figure 1. Importantly, when the exponential term does not decay fast enough or is not present at all (see all traces in Figure 1 for which $r < 1.007$), the force-indentation data remain aperiodic and data spectra computed with the DFT would not correspond to physical behavior. With the introduction of the exponential term in equation 5, it is easy to view the Z-transform and modified Fourier transform in terms of a conventional Fourier transform. We have found that it is most convenient to perform these transforms using traditional fast Fourier transform algorithms and values of $r$ greater than 1.

$$Z\{X\} = \sum_{n=0}^{\infty} x[n]z^{-n} = \sum_{n=0}^{\infty} x[n]r^{-n}e^{-i\omega n} = \sum_{n=0}^{\infty} (x[n]r^{-n})e^{-i\omega n} \tag{5}$$

$$MDFT\{X, r_0\} = \sum_{n=0}^{\infty} x[n]r_0^{-n}e^{-i\omega n} = FFT\{x[n]r_0^{-n}\} \tag{6}$$

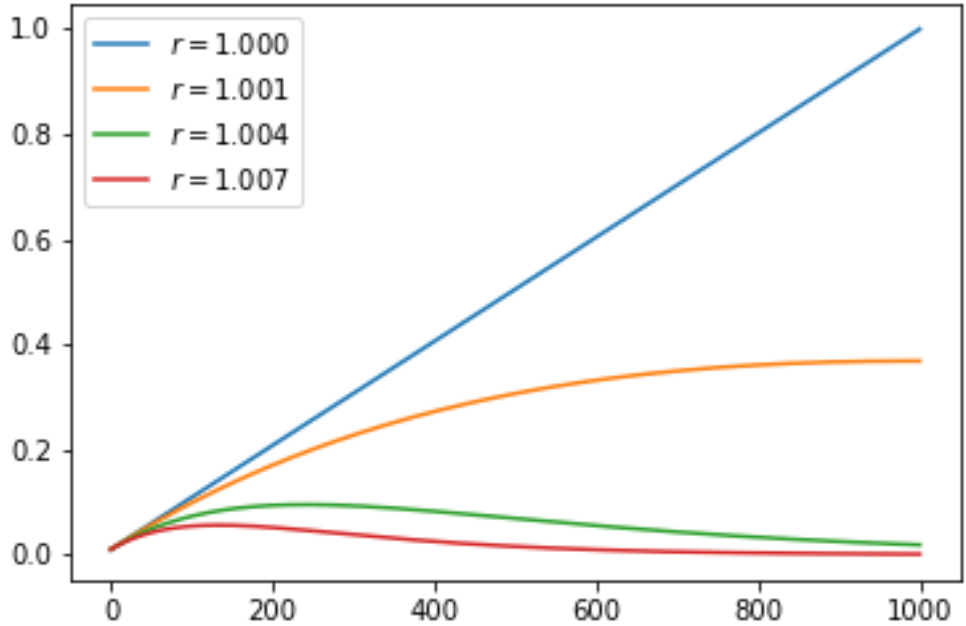

**Figure 1** An arbitrary linear signal with different degrees of binding through the exponential term $r^{-n}$, corresponding to different circles on the Z-domain.

Through selection criteria which will be outlined in section III.b, one should find a value for the constant $r$, termed the decay constant, which allows for the simultaneous calculation of the MDFT of both the force and indentation signals. Once this has been done, one can simply divide the two spectra in the complex domain, making sure to include the proper constant for the contact mechanics geometry under consideration, which directly leads to the relaxance or retardance of the material, as seen in Equation 7. It is important to note that the real and imaginary components of these quantities do not directly correspond to the storage and loss behavior of the material, since they are not defined in the Fourier domain (i.e., since $r$ is not equal to 1). Although they may be numerically close, the Fourier domain behavior needs to be estimated from this modified Fourier domain behavior by fitting a viscoelastic model that is defined in the entire Z-domain. The resulting fit can then be expressed setting $r$ to be equal to 1, to give an estimation of the storage and loss behavior. [9]

$$\frac{MDFT\{F, r_0\}}{\frac{16\sqrt{R}}{3} MDFT\left\{h^{\frac{3}{2}}, r_0\right\}} = Q(\omega, r_0) = \frac{1}{U(\omega, r_0)} \tag{7}$$

### III. METHOD IN PRACTICE

#### a. OUTLINE

In this section, we discuss how to implement the previously outlined procedure in practice. A flow chart outlining the process can be found in Figure 2. The process steps can be summarized as i) obtaining stress and strain from the AFM force-indentation data using contact mechanics, ii) recommended denoising of the data, iii) estimating a suitable decay constant for the MDFT, iv) calculating the MDFT of the stress and strain (or force and indentation), v) averaging several spectra to further reduce the noise, vi) calculating the relaxance or retardance, and finally, vii) obtaining the modified Fourier domain forms of the storage and loss behavior of the material. After obtaining the relaxance or retardance, one may also fit the spectra using a viscoelastic model of choice to obtain an estimation for the true storage and loss behavior. Having access to a frequency representation of the material behavior greatly assists in the proper selection of the viscoelastic model.

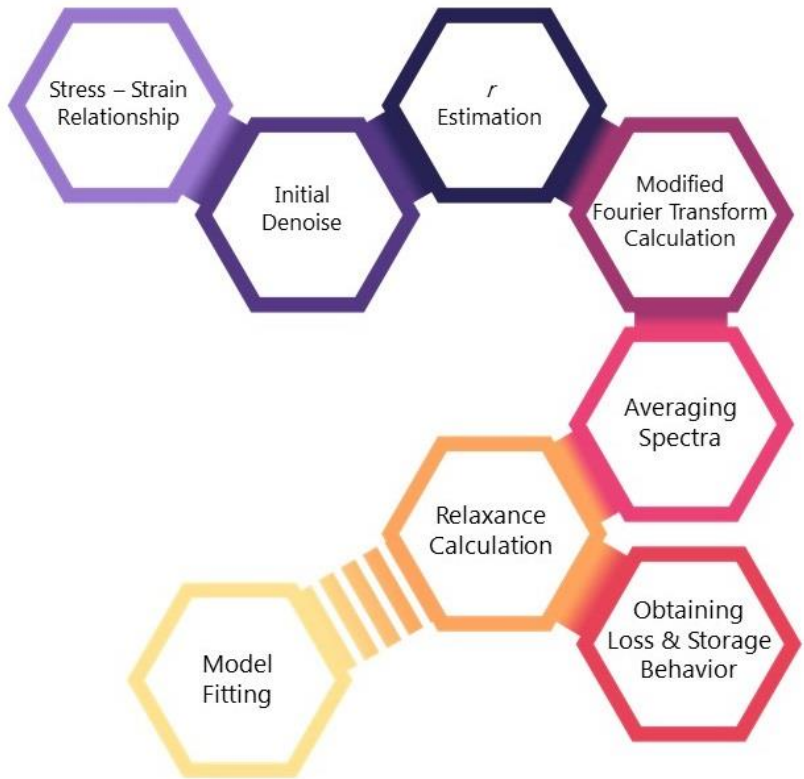

**Figure 2** A flowchart summarizing the proposed technique. After calculation of the modified Fourier spectrum of the stress and strain (or force and indentation), the data is handled in the complex domain rather than the time domain. Before calculating the relaxance of the material, averaging the stress and strain (or force and indentation) spectra among themselves is important to prevent unrealistic local peaks in the relaxance when plotted vs. frequency. This is also important to avoid division by zero in the calculations. Once the relaxance has been calculated, the user may decide to fit it to a viscoelastic model of choice.

We now elaborate on each step in the process using a set of example force curves obtained for a nylon sample cut from a rod (Figure 3A). The force-distance curves, which are plotted in Figure 3B, were obtained with an AFM cantilever having a 0.5 N/m stiffness and 35 nm tip radius, and using a 250 nm/s approach speed and a 50 kHz sampling rate. Figure 3B consists of 22 force-distance curves plotted together. As is evident in the figure, the raw force-indentation curves contain a considerable amount of noise. Before calculating the modified spectrum for the curves, we have discarded the non-contact portion of the curves, keeping only the repulsive portion of the data, and have denoised the curves using a moving average with a window size of 8% of the total data length of the respective curve. In Figure 3C, a single representative raw FD curve and the denoised curve with the moving average correction can be seen.

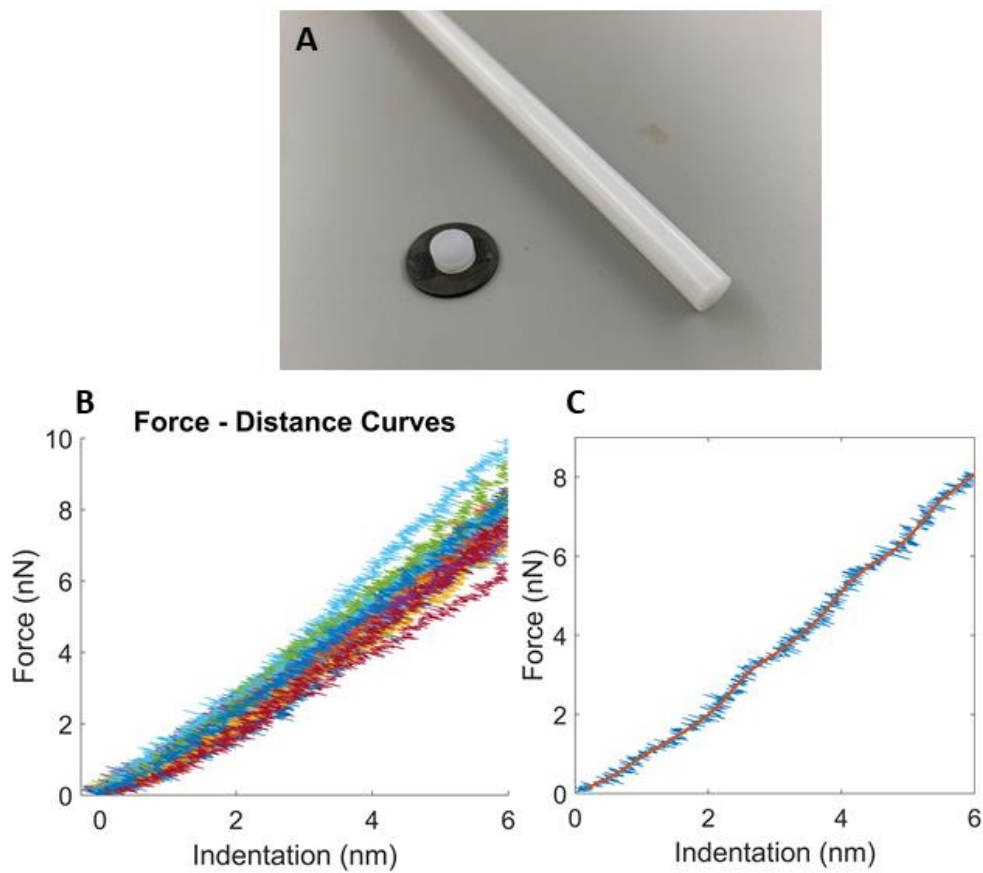

**Figure 3** A) Nylon sample cut from a rod; B) collection of 22 Force-Distance curves obtained on the nylon sample; C) one representative Force-Distance curve before and after applying a moving average with a window size of 8% of the total data length of the respective curve in order to denoise it.

b. **ESTIMATION OF THE DECAY CONSTANT**

As discussed in the theory section, to calculate the MDFT of the force and indentation signals, one needs a suitably large decay constant $r$ to bind the signals within the time window of measurement, in order to obtain physically meaningful results. [9] On the other hand, by using too large a decay constant, one significantly suppresses the signal which may result in too much valuable information being lost. Therefore, one needs a value that is large enough to bind the signals, but small enough to preserve as much fidelity as possible to the original, true behavior. In this work, we have used an algorithm that selects an appropriate time coefficient, such that the endpoint of the bound signal is equal to its first non-zero element at the beginning of the sequence. This algorithm is further outlined in the Supporting Information. Although the algorithm gives acceptable decay constants, it does not necessarily provide 'the best time constant. In order to get a cleaner-looking spectrum using this method, one may still need to adjust the resulting time constant within an order of magnitude (at most).

Since one needs to calculate the quotient of the force and indentation spectra to obtain the relaxance or retardance, we argue that such spectra should be defined on similar regions of the Z-domain. Thus, the decay constant should be the same for the force and the indentation. Since the decay constant needs to be able to bind both signals, it is prudent to use the larger of the two decay constants calculated for the force and indentation. Likewise, when dealing with multiple force-distance curves, one should use the same decay constant for all curves. Different time coefficients would project the signals onto different circles in the z-plane. When averaging curves or comparing them to each other, it would be inappropriate to compare across different circles.

Comparing across experiments with different sampling frequencies, similarly careful treatment is needed. In the previous treatments, the progressions of the signals were not directly related to time; rather, they were related to the index of the signal. However, it makes most physical sense to consider the force and indentation as signals in time. This is most conveniently done by tying the index value to the sampling time through the product $t = \Delta t\, n$. Likewise, the decay constant $r$ can be defined as in Equation 8 below. Experimental results obtained with different sampling frequencies should then be compared under identical values of $r_t$ rather than $r$.

$$r^{-n} = r^{-\frac{t}{\Delta t}} = r_t^{-t} \tag{8}$$

### c. CALCULATION OF THE TRANSFORMED SIGNALS

After obtaining a suitable time constant, one can transform the force and indentation data sequences into the modified Fourier domain. One may notice that even after initial denoising in the time domain, the force and indentation spectra may still contain considerable noise. As this noise may cause a few extreme outlying points, their effect may be heavily amplified after dividing the spectra to obtain the retardance or relaxance. Due to machine precision limitations, this issue may not be fixable with a simple denoising scheme. Thus, separately obtaining an average spectrum for all force signals and all indentation signals before performing the division is most ideal in order to reduce the effects of noise. Averaging spectra from multiple experiments is a common technique in many different spectroscopy techniques that utilize the Fourier transform, such as Raman spectroscopy, FTIR, AFM thermal tuning, etc.[8,22–24] Furthermore, if desired, averaging curves obtained from different locations on the sample may help to obtain an average material characteristic as opposed to local characteristics. Averaging spectra obtained from different signals with the same length is quite trivial. However, the force and indentation signals we obtain from different experiments may have slightly different numbers of data points. In order to be able to average them on the same frequencies, one should calculate the modified Fourier transform with a set transform length given by the length of the shortest experimental signal.

After obtaining the averaged force and indentation spectra, the relaxance or retardance of the material can be calculated using the relations described in Equation 9. Again, taking the real and imaginary components of the relaxance or retardance gives the modified storage and loss behaviors, respectively, which are numerically close but not exactly equal to the true storage and loss behaviors. [9] One can use this modified relaxance or retardance as a transfer function to determine material responses given a specific input, but in order to obtain the true storage and loss behavior, it is necessary to utilize a viscoelastic model defined in the Z-domain to convert between the modified Fourier and the Fourier domains.

### d. ANALYTICAL MODELS IN THE Z-DOMAIN

Viscoelastic models can provide a good understanding of how materials behave and are a convenient way to represent their properties. In this work, we also use them as bridges between MDFT storage or loss behavior and storage or loss moduli. In this section, we provide mathematical expressions for four commonly used viscoelastic models in the Z-domain.

The Generalized Maxwell (Maxwell-Wiechert) and Generalized Kelvin-Voigt models are commonly used to describe the relaxance and retardance through an equivalent spring-dashpot model, as seen in Equations 9 and 10. It is noteworthy to mention that these two models are congruent. For each Generalized Maxwell model an equivalent Generalized Kelvin-Voigt model can be constructed and vice versa.[25] Here, $\Delta t$ is the sampling timestep of the experiment and $N$ is the number of Maxwell "arms" of the model, which may be increased so that the experimental data is better fit by the model; however, this overfitting is not necessarily advisable. From a mathematical perspective, $N$ determines the number of peaks in the loss modulus or compliance. [9] Therefore, we argue that one should select $N$ to be equal to the number of distinct peaks in the imaginary component of the experimentally obtained relaxance or retardance, even in cases where increasing $N$ further would give a better numerical fit. For instance, the data seen in Figure 4 indicates that $N$ should be either 1 or 2.

$$Q_{Maxwell} = G_g - \sum_i^N \frac{G_i}{1 + \frac{\tau_i}{\Delta t}\left(1 - \frac{e^{-i\omega}}{r}\right)} \tag{9}$$

$$U_{Voigt} = J_g + \sum_i^N \frac{J_i}{1 + \frac{\tau_i}{\Delta t}\left(1 - \frac{e^{-i\omega}}{r}\right)} \tag{10}$$

For some materials, it may be desirable to use a different definition of the material behavior that may not be possible to accomplish with one of the above models. [2,5,6] In this case, one may use fractional viscoelastic elements, referred to as a spring-pots (S-P), or a power law (PLR) model, for example. Both approaches are defined in Equations 11 and 12. In the case of the spring-pot model, the $\beta$ term represents the fractional order of the differential stress-strain relationship, whereby its values range between 0 and 1, corresponding to pure elastic and pure fluidic behavior, respectively.[6] The $C_\beta$ term is a simple scaling constant that has units of $\frac{Pa}{s^\beta}$. The power law model behaves in a very similar manner to the spring-pot in that it contains a scaling constant, $E_0$, with units of Pa, and an exponential term, $n$, which ranges from 0 to 1, exclusive.[5,6]

$$Q_{S-P} = \frac{1}{U_{SP}} = C_\beta \left(1 - \frac{e^{-i\omega}}{r}\right)^\beta \tag{11}$$

$$Q_{PLR} = E_0 \left(\frac{t_0}{\Delta t}\right)^n \Gamma(1-n) \left(1 - \frac{e^{-i\omega}}{r}\right)^n \tag{12}$$

Aside from converting between the Fourier and modified Fourier domains, model fits are desirable by many practitioners in order to obtain a parameterized description of the material, which readily yields quantities such as the relaxation time, for example. One advantage of obtaining the frequency domain behavior of a material directly from an experiment first, without fitting, is that one can then visually assess exactly what type of material model would be best suited to describe the material being analyzed. For example, it can be seen from Figure 4 that the nylon sample behavior agrees well with a single-arm Maxwell (Standard Linear Solid) model. Figure 5 provides estimates of the true storage and loss moduli based on this model.

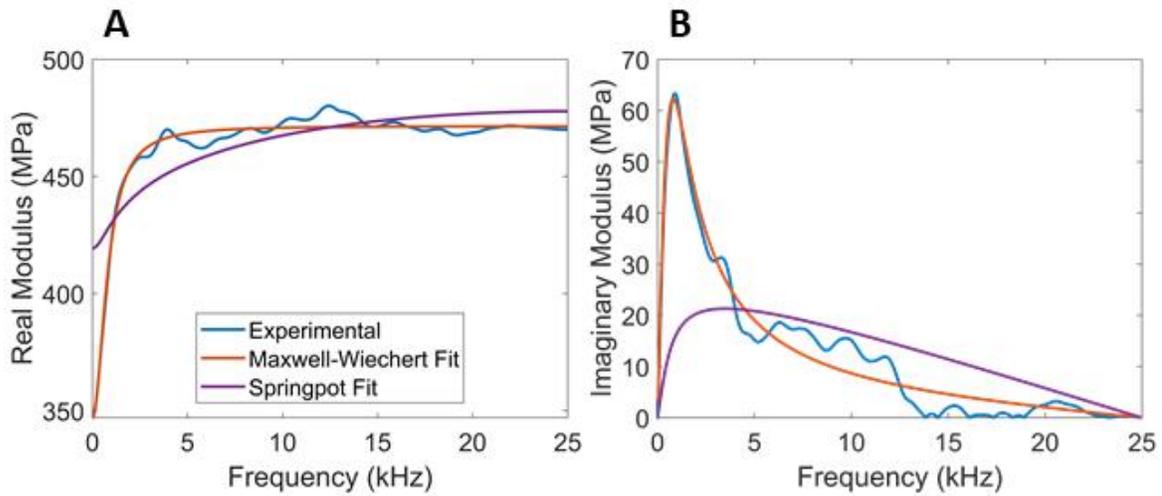

**Figure 4** A) Real and B) imaginary parts of the experimentally obtained, frequency averaged relaxance of the nylon sample from Figure 3, as well as the model fits using the Maxwell-Wiechert and Spring-pot models. The fit to the Maxwell-Wiechert model is particularly good. Different degrees of agreement may be observed for different viscoelastic models when studying different materials. One significant advantage of having access to the model-free frequency behavior is to be able to compare model and material behavior before choosing a particular viscoelastic model.

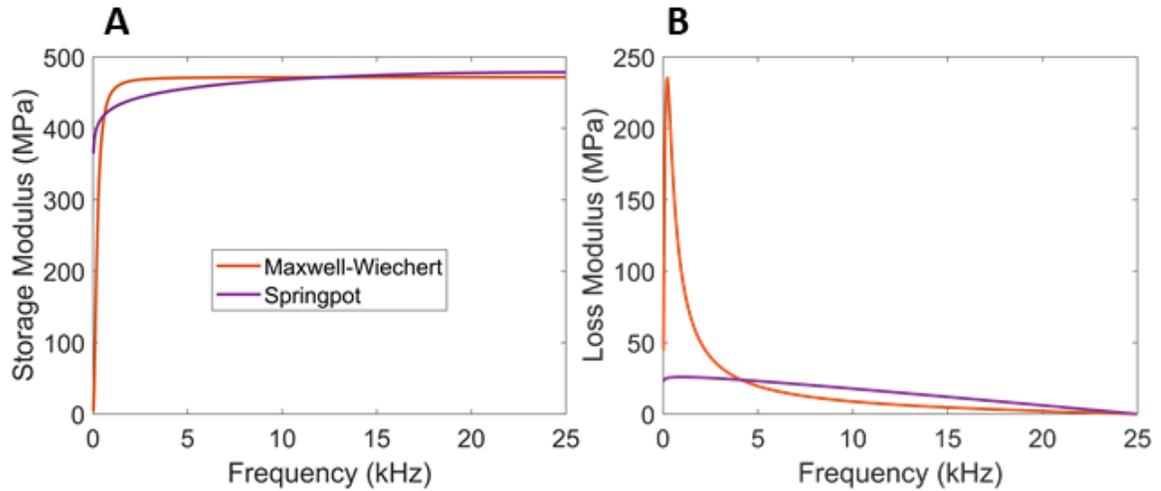

**Figure 5** A) Storage and B) loss modulus obtained from the models fitted to the data in Figure 4. Note that the moduli tend towards very small values at low frequency. Such a phenomenon is known as creep.

The proposed method is quite versatile and could also be applied to non-polymeric materials, and even to arbitrary composite samples whose molecular structure is not known in detail. In order to demonstrate this, we have repeated the same procedure for a freshly cut orange peel sample (Figure S3), using an AFM cantilever with 0.1 N/m stiffness and 10 µm spherical colloidal bead probe. The experimental results and model fits can be seen in Figure 6. Clearly, the results do not correspond to the behavior of a single, specific material but such an analysis may still be quite useful, for example, for quality control purposes in industrial settings, among other applications.

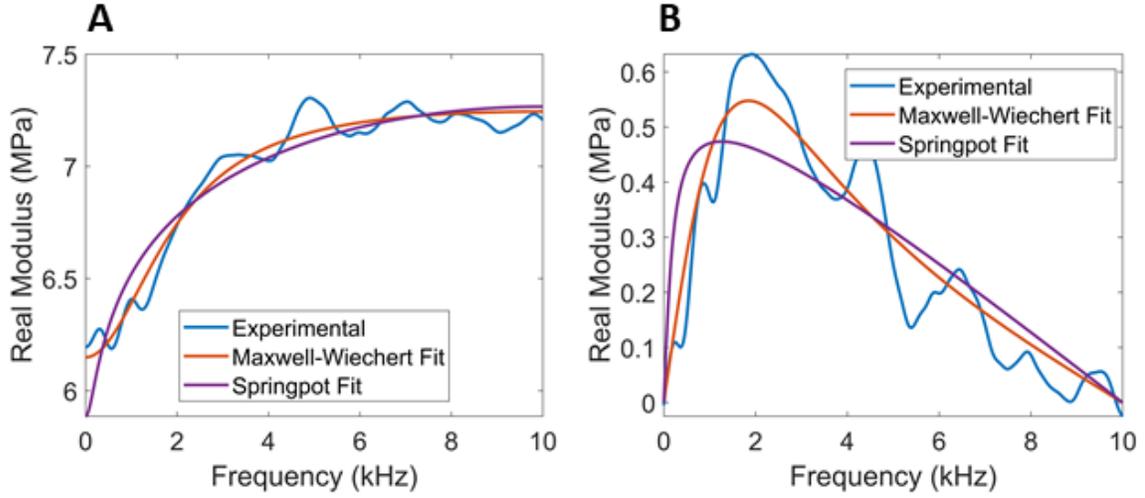

**Figure 6** A) Storage and B) loss behaviors obtained for an orange peel sample.

In this manuscript, we have discussed the method from an AFM point of view, but the method is also applicable to other experimental techniques that record the stress-strain relationship of a material, such as macroscale compressive and tensile testing instruments. Since these systems generally provide stress and strain outputs rather than a force–indentation curve, we would base our analysis on Equation 13, which is derived from Equation 1, and we would follow the same steps to characterize the material as we have done in this manuscript.

$$\sigma(\omega, r_0) = Q(\omega, r_0)\, \epsilon(\omega, r_0) \tag{13}$$

IV.        CONCLUSION

We have demonstrated a method to obtain the viscoelastic properties of a material in the modified Fourier domain using data from force-distance atomic force microscopy experiments. We have also provided a means to convert the results from the modified Fourier domain to the Fourier domain in order to obtain approximations of the storage and loss behavior of the material. The proposed method utilizes the modified discrete Fourier transform (MDFT). Unlike traditional Fourier transform techniques, the MDFT has the added benefit of being able to handle the numerically unbounded (aperiodic) inputs obtained from the AFM force-distance experiments. This method also serves as a useful tool for classical viscoelastic model fitting.

**ACKNOWLEDGEMENT**

B.U. gratefully acknowledges support from the US Department of Energy, Office of Science, Basic Energy Sciences, under Award No. DE-SC0018041. M.R.M and S.D.S. gratefully acknowledge support from the US National Science Foundation, under award CMMI-2019507.

# Direct Measurement of Storage and Loss Behavior in AFM Force-Distance Experiments Using the Modified Fourier Transformation

*Berkin Uluutku, Marshall R. McCraw and Santiago D. Solares\**

*Department of Mechanical and Aerospace Engineering, The George Washington University School of Engineering and Applied Science, Washington, District of Columbia, USA*

*Corresponding Author: Santiago D. Solares – ssolares@gwu.edu

## I. SELECTION OF A STABLE DECAY CONSTANT

When calculating the modified discrete Fourier transform (MDFT) of a numerically unbounded signal, one needs to select a time decay constant that is large enough to bind the signal and avoid spectral leakage, but small enough to preserve as much as possible of the fidelity of the bounded signal to its original form. We accomplish this by solving for the decay constant that makes the signal periodic, hence the first non-zero value of the damped signal is equal to its last value, as seen in equation S1 and represented visually in figure S1. Isolating for the decay constant $r$ yields equation S2. Although this method gives repeatable results, the value of the decay constant may need to be adjusted within an order of magnitude in some cases. Based on our experience, decay constants that are too small may result in overly noisy spectra, and unphysicsal behavior at frequency extremes. Likewise, an unexpectedly smooth spectra from an experimental data may mean the decay constant is larger than needed.

$$x[n_0]r^{-n_0} = x[N]r^{-N} \tag{S1}$$

$$r = \left(\frac{x[N]}{x[n_0]}\right)^{\frac{1}{N-n_0}} \tag{S2}$$

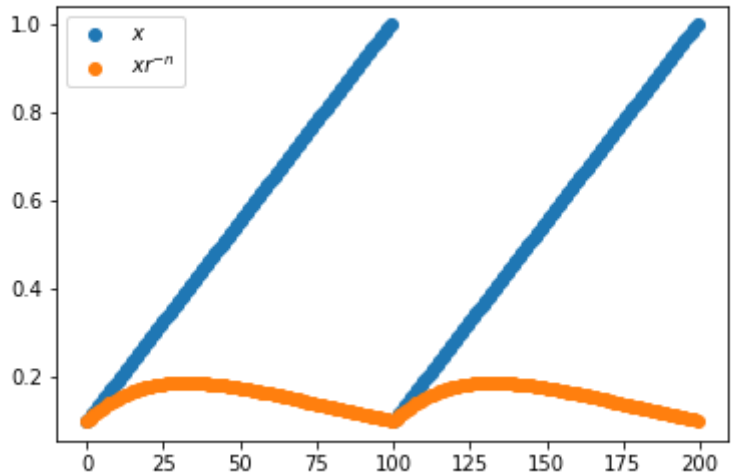

**Figure S1** The undamped (unbounded) signal (blue) will not smoothly repeat and will therefore have spectral leakage when its Fourier transform is taken. The same signal is then damped using a decay constant that matches its starting and ending points (orange), which gives a smoothly repeating signal having a physically meaningful Fourier transform.

When selecting an ideal decay constant, one should expect that for a number of force-distance curves from the same experiment, the decay constants should be similar. This is illustrated by the histogram shown in figure S2. Any outlying decay constants could be associated with noisy or erroneous data. This analysis is useful when analyzing large numbers of force curves, whereby one would discard data that has a decay constant differing significantly from the mean decay constant of the data set. This is especially advantageous when applying the averaging scheme discussed in the main paper.

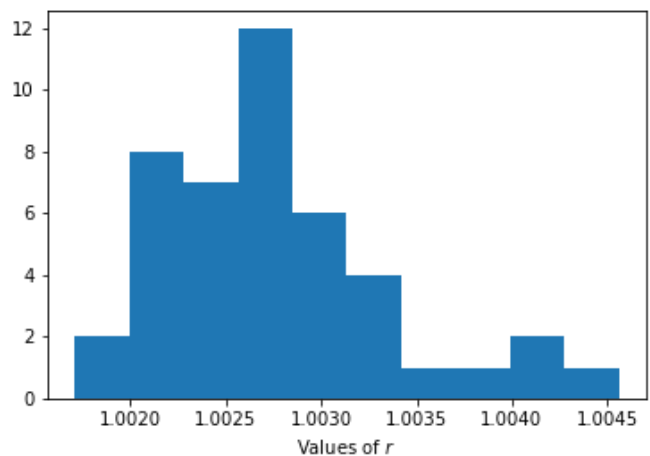

**Figure S2** A histogram of the decay constants for the force-distance curves provided in Figure 3 of the main paper. Note how the various decay constants do not significantly differ in value, as the force curves themselves are quite similarly shaped.

## II. SUGGESTED METHOD OF TAKING MODIFIED DISCRETE FOURIER TRANSFORM

As discussed in the main paper, the Z-transform, discrete Fourier transform, and modified discrete Fourier transform (MDFT) are closely related.[1] As seen in Table S1, the MDFT is obtained by evaluating the Z-transform along a fixed radius in the Z-domain. If that radius is set to 1, one obtains the discrete Fourier transform. Furthermore, we can see that the MDFT is also obtained by taking the discrete Fourier transform of the product of the signal $x[n]$ and the decaying exponential $r^{-n}$. This allows the use of traditional fast Fourier transform algorithms when working with the proposed MDFT methods. Additionally, since the best results are obtained when averaging multiple curves together, we recommend using an "N-point Fourier transform" and setting N to be equal to the shortest curve length in the dataset.

**Table S1** Discrete transforms and their equations.

| Transform Name | Equation | Transform Variables |
|---|---|---|
| Z | $\sum_{n=0}^{\infty} x[n] r^{-n} e^{-i\omega n}$ | $r, \omega$ |
| Discrete Fourier | $\sum_{n=0}^{\infty} x[n] e^{-i\omega n}$ | $\omega$ |
| Discrete Modified Fourier | $\sum_{n=0}^{\infty} x[n] r^{-n} e^{-i\omega n}$ | $\omega$ |

## III. DERIVATION OF MODELS

To derive the relaxance of a Generalized Maxwell model in the Z-domain, we start with the definition of the derivative in the Z-domain, which is given in equation S3.[2] We can then write the stress-strain relationship for a single arm of the Maxwell model (a spring and dashpot in series) (equation S4), and subsequently represent it in the Z-domain (equation S5). The individual spring element can then be treated in a similar fashion, as described in equations S6 and S7. The relaxance of the model is then obtained in equation S8 by summing the stresses in each arm and dividing the time constant $\tau$ by $\Delta t$ to represent it as an index rather than in units of time.

$$Z\left\{\frac{df(t)}{dt}\right\} = (1 - z^{-1})Z\{f(t)\} = \left(1 - \frac{e^{-iw}}{r}\right)Z\{f(t)\} \tag{S3}$$

$$\dot{\varepsilon} = \frac{\dot{\sigma}}{G_i} + \frac{\sigma}{G_i \tau_i} \tag{S4}$$

$$\sigma(z) = \varepsilon(z) \frac{(1 - z^{-1})}{\frac{(1 - z^{-1})}{G_i} + \frac{1}{G_i \tau_i}} \tag{S5}$$

$$\sigma = G_e \, \varepsilon \tag{S6}$$

$$\sigma(z) = G_e \, \varepsilon(z) \tag{S7}$$

$$Q_{Maxwell}(z) = \frac{\sigma(z)}{\varepsilon(z)} = G_g - \sum_i^N \frac{G_i}{1 + \frac{\tau_i}{\Delta t}\left(1 - \frac{e^{-i\omega}}{r}\right)} \tag{S8}$$

Following the same procedure as above, one can similarly derive the retardance of a Generalized Kelvin-Voigt model.[2] We start with the stress-strain relation for a single Kelvin-Voigt unit, consisting of a spring and a dashpot in parallel, as seen in equation S9. By transforming this equation into the Z-domain, one obtains equation S10. Then the stress-strain relationship of the single spring element is given in equation S11, which is then transformed into the Z-domain, as seen in equation S12. By summing over the strains in every unit, the retardance is obtained as in equation S13. Similar to the relaxation time above, the retardation time $\tau$ is normalized by $\Delta t$ to keep it in terms of the index of the signal and not in units of time.

$$\sigma = \frac{\varepsilon}{J_i} + \frac{\tau_i}{J_i}\dot{\varepsilon} \tag{S9}$$

$$\varepsilon(z) = \sigma(z)\frac{J_i}{1 + \tau_i(1 - z^{-1})} \tag{S10}$$

$$\varepsilon = J_g\,\sigma \tag{S11}$$

$$\varepsilon(z) = J_g\,\sigma(z) \tag{S12}$$

$$U_{Voigt}(z) = \frac{\varepsilon(z)}{\sigma(z)} = J_g + \sum_i^N \frac{J_i}{1 + \frac{\tau_i}{\Delta t}\left(1 - \frac{e^{-i\omega}}{r}\right)} \tag{S13}$$

The relaxance of the spring-pot model is straightforward. Taking the Z-transform of the fractional-order derivative in equation S14 yields equation S15.[3,4] Rearranging in terms of relaxance and substituting for polar coordinates yields the form seen in equation S16.

$$\sigma = C_\beta \frac{d^\beta \varepsilon}{dt^\beta} \tag{S14}$$

$$\sigma(z) = C_\beta\,(1 - z^{-1})^\beta\,\varepsilon(z) \tag{S15}$$

$$Q_{S-P} = C_\beta \left(1 - \frac{e^{-i\omega}}{r}\right)^\beta \tag{S16}$$

To derive the relaxance of the power-law model, one needs to consider the definition of the power-law relaxation modulus (Equation S17).[3,4] More specifically, the relaxation modulus $G$ is equal to the time derivative of the relaxance $Q$ of the material. We start by taking the Z-transform of the power-law relaxation modulus in equation S18. The result needs to be corrected to account for $n$ not being an integer, using the identity given in equation S20. The relaxance can then be obtained by taking the derivative of the relaxation modulus. The final result is shown in equation S23, where the Z-domain variable has been represented in polar coordinates and the time constant $t_0$ has been normalized by the timestep $\Delta t$ to make it an index rather than a unit of time.

$$Q = \frac{d}{dt} G \tag{S17}$$

$$G_{PLR} = E_0 \left(\frac{t}{t_0}\right)^{-n} \tag{S18}$$

$$G_{PLR}(z) = E_0 \, t_0^n \, (-n)! \, (1 - z^{-1})^{n-1} \tag{S19}$$

$$\Gamma(x) = (x - 1)! \tag{S20}$$

$$G_{PLR}(z) = E_0 \, t_0^n \, \Gamma(1 - n) \, (1 - z^{-1})^{n-1} \tag{S21}$$

$$Q_{PLR}(z) = (1 - z^{-1}) \, E_0 \, t_0^n \, \Gamma(1 - n) \, (1 - z^{-1})^{n-1} \tag{S22}$$

$$Q_{PLR}(z) = E_0 \left(\frac{t_0}{\Delta t}\right)^n \Gamma(1 - n) \left(1 - \frac{e^{-i\omega}}{r}\right)^n \tag{S23}$$

An astute reader might notice that the power-law and the spring-pot models are equivalent through the conversion seen in equation S24. This relationship could be used to compare materials characterized using different models.

$$C_\beta = E_0 \left(\frac{t_0}{\Delta t}\right)^n \Gamma(1 - n) \tag{S24}$$

IV.     DIFFERENT CONTACT MECHANICS GEOMETRY

One may wish to use the proposed method to analyze data obtained with AFM tip geometries other than a spherical probe.  To consider such cases we start with the Lee and Radok model for viscoelastic contact as seen in equation S25.[5] Instead of leaving explicit geometry parameters in this equation, we have kept it in terms of the two general contact geometry parameters $\alpha$ and $\beta$. Taking the modified discrete Fourier transform of this equation gives the form of the relaxance and retardance provided in equation S26. The values of these parameters are listed in table S2 for spherical, conical, and flat punch contact geometries, where $v$ is the Poisson's ratio of the material, typically taken as 0.5 (for incompressible materials).

$$F = \alpha \int_0^{t_f} Q(t-u)\, h^\beta(u)\, du \tag{S25}$$

$$\frac{MDFT\{F, r_0\}}{\alpha\, MDFT\{h^\beta, r_0\}} = Q(\omega, r_0) = \frac{1}{U(\omega, r_0)} \tag{S26}$$

**Table S2** Parameters for various contact geometries.

| Contact Type | $\alpha$ | $\beta$ |
|---|---|---|
| Spherical | $\dfrac{4\sqrt{r}}{3(1-v^2)}$ | $\dfrac{3}{2}$ |
| Conical | $\dfrac{2\tan(\theta)}{\pi(1-v^2)}$ | $2$ |
| Flat Punch | $\dfrac{2R}{(1-v^2)}$ | $1$ |

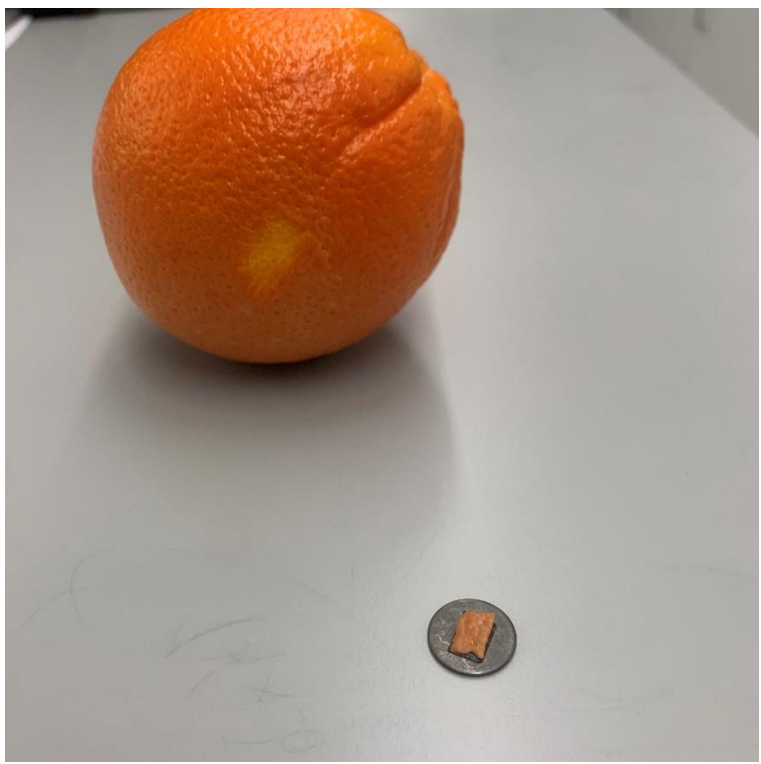

**Figure S3** A representative image of the orange peel sample used to construct Figure 6 of the main manuscript. After acquiring a cut from the peel using a scalpel, the peel was taped to a metal sample holder and force curves were obtained.